\documentstyle[aps,prl,twocolumn,epsfig]{revtex}
\begin{document}

%
\twocolumn[\hsize\textwidth\columnwidth\hsize\csname@twocolumnfalse%
\endcsname

\title {Superdiffusion and Out-of-equilibrium Chaotic Dynamics 
with Many Degrees of Freedoms}

\author{   Vito Latora $^{(a),(1)}$   }

\address{Center for Theoretical Physics, Laboratory for Nuclear Sciences 
and Department of Physics, 
\\ Massachusetts Institute of Technology, Cambridge, Massachusetts 02139, USA
\\and 
\\Department  of Physics, Harvard University, 
Cambridge, Massachusetts 02138, USA}

\author{  Andrea Rapisarda $^{(b),(2)}$ }

\address{ Dipartimento di Fisica, Universit\'a di Catania \\ 
and Istituto Nazionale di Fisica Nucleare, Sezione di Catania
\\
Corso Italia 57, I-95129 Catania, Italy } 

\author{   Stefano Ruffo  $^{(c),(3)}$  }

\address{Dipartimento di Energetica, Universit\'a di Firenze, INFM  and INFN
Via S. Marta 3, Firenze, Italy}

\date{26 April 1999}

\bigskip
\centerline {Revised version accepted for publication by Physical Review Letters, scheduled for 
september (1999)}
\bigskip
 
\maketitle


\begin{abstract}

We study the link between 
relaxation to the equilibrium and anomalous superdiffusive
motion in a classical N-body hamiltonian system with long-range 
interaction showing a second-order phase-transition in the
canonical ensemble. Anomalous diffusion is 
observed only in a transient out-of-equilibrium regime and for a small range 
of energy, below the critical one.
Superdiffusion is due to L\'evy walks of single particles and is checked
independently through the second moment of the distribution, power spectra,
trapping and
 walking time probabilities.
Diffusion becomes normal at equilibrium, after a relaxation
time which diverges with N. 

\end{abstract}


]

In the recent years there has been an increasing 
interest for physical phenomena 
which violate the central limit theorem such as anomalous diffusion 
and L\'evy walks. These violations are not an exception
in Nature and have been observed in many different fields
and also in connection with deterministic chaos in low dimensional 
systems \cite{geisel,sol,tsallis,kla,levypro}.
The availability of more powerful computers 
has made possible to study deterministic chaos and subdiffusive
motion in systems with many degrees of freedom using nearest-neighbour 
coupled symplectic maps~\cite{kaneko}. 
In a very recent work superdiffusive motion has been 
found in a 
 N-body Hamiltonian system 
with  long-range 
couplings~\cite{torc}.
The mechanism underlying this anomalous diffusion is similar
to the one proposed by Geisel\cite{geisel} in ``egg-crate"
two-dimensional potentials.

In this Letter we present a novel study of superdiffusion 
and L\'evy walks in  a
 Hamiltonian system of N fully coupled rotors
(called Hamiltonian Mean Field, HMF)
\cite{antoni,latora}.
The new interesting result is that, in HMF superdiffusion is
connected to  the presence of
quasi-stationary non-equilibrium states, rather than to the
mechanism proposed by Geisel\cite{geisel} and found also in \cite{torc}.
HMF has been used to investigate relaxation to 
thermodynamical equilibrium for systems with 
long-range interactions.
It has been studied both at a macroscopic level, 
by means of the canonical formalism and at a 
microscopic dynamical level. 
The canonical ensemble predicts a second-order 
phase transition from a clustered phase to a 
homogeneous one~\cite{antoni,latora}. 
On the other hand, microcanonical simulations show a strong 
chaotic behavior in the region below the critical energy;   
Lyapunov Exponents and Kolmogorov-Sinai entropy reach a maximum at 
the critical point~\cite{latora}.
These results have been confirmed also for long but finite-range
interactions~\cite{celia}.
Of particular importance for this Letter are the 
results obtained in Ref.~\cite{latora} concerning the  
discrepancies between microcanonical results and  
canonical predictions. 
In fact, numerical simulations performed at constant 
energy reveal the existence of out-of-equilibrium 
Quasi-Stationary States (QSS) with an extremely slow 
relaxation to equilibrium.  
In Ref.~\cite{ruffo} these QSS are shown to become 
stationary solutions in the continuum limit.

The main results of this letter are:

1) we find evidence of an anomalous superdiffusive behavior 
below the critical energy. Anomalous diffusion changes 
to normal one after a crossover time $\tau_c$, as also found by other
%
authors~\cite{geisel,sol,kaneko,torc,grigo,barkai}. 
%
Power spectra
confirm the presence of the anomaly; 

2) the superdiffusive behavior is connected to the presence of 
out-of-equilibrium QSS . 
We give substantial numerical evidence that the crossover time  $\tau_c$ 
coincides with $\tau_r$, the time needed for QSS to relax to 
canonical equilibrium;

3) we give an interpretation of our results in terms of 
L\'evy walks, that are originated by chaotic transport of 
each rotor,  
which moves with an energy not constant in time and alternately sticks to
the cluster, which has a quasi-regular motion
 or undergoes free walks far from it with a 
constant velocity much greater  than that of the cluster.
%
 Trapping time and walking 
time probability distributions show a power law behavior. The 
corresponding exponents can be related to the superdiffusion exponent
using the model of Ref.\cite{kla} 
and are very similar   to those found in the fluid flow experiment
of Solomon et al. \cite{sol}.

In the following we remind the formalism and then we 
review the numerical results.
HMF describes a system of $N$ classical particles (or rotors) 
characterized by the angles $\theta_i$ and the conjugate momenta 
$p_i$. 
Each rotor interact with all the others 
according to the following Hamiltonian:  
\begin{equation}
        H(\theta,p)=K+V ~,
\end{equation}
where 
\begin{equation}
       K= \sum_{i=1}^N  \frac{{p_i}^2}{2} ~~~~~ 
       V= \frac{1}{2N} \sum_{i,j=1}^N  [1-cos(\theta_i -\theta_j)]
\end{equation}
are the kinetic and potential energy. 
One can define a spin vector associated to each rotor 
${\bf m}_i=[cos(\theta_i), sin(\theta_i)]$ 
and a total magnetization 
${\bf M}={\frac{1}{N}}\sum_{i=1}^N {\bf m}_i$. 
The Hamiltonian then describes N classical 
spins similarly to the XY model.  
This system has a ferromagnetic 
second-order phase transition from a clustered phase 
to a homogeneous one at a critical temperature $T_c=0.5$ 
and a corresponding critical energy $U_c=E_c/N=0.75$ 
(see ref.\cite{antoni,latora}).
The equations of motion for the $N$
rotors are given by: 
\begin{equation} 
\dot{r_i}={p_i}, ~~~\dot{p_i}  = - M sin(\theta_i - \phi ) ~~~,~~~ 
i=1,...,N~~~,
\label{eqmoto} 
\end{equation}
where $(M,\phi)$ are respectively modulus and phase 
of the total magnetization vector $\bf M $. 
These equations are formally equivalent to those of
a perturbed pendulum.                                   
To study relaxation to canonical equilibrium, 
we solve these equations on the computer  
using fourth order symplectic algorithms 
(the details can be found in ref. \cite{latora}).  
We start the system in a given initial distribution 
and we compute ${\theta_i, p_i}$ at each time step, and from them   
the total magnetization $M$ and temperature $T$
(through the relation $T=2<K>/N$).  
We consider systems with an increasing size $N$ and different 
energies $U=E/N$.

Diffusion and transport of a particle in a medium or in a fluid 
flow are characterized by the average square displacement $\sigma^2(t)$
in the long-time limit.  
In general, one has 
\begin{equation}
\label{anoma}
    \sigma^2(t) \sim  t^{\alpha}
\end{equation}
with $\alpha=1$ for normal diffusion.   
All the processes with $\alpha \ne 1$ are termed anomalous 
diffusion, namely subdiffusion for $0<\alpha<1$ and 
superdiffusion for $1<\alpha<2$. 

In order to study anomalous diffusion in HMF we 
follow the dynamics of N rotors starting the system in a 
``water bag'', i.e. a far-off-equilibrium initial condition
obtained by putting all the rotors at $\theta_i=0$ and
giving them a uniform distribution of momenta with a
finite width centered around zero. 
We compute the variance of the one-particle angle $\theta$ 
according to the expression
\begin{equation}
    \sigma_{\theta}^2(t) = < (\theta - <\theta> )^2 >
~~~~~~~~,
\end{equation}
where  $< ~.~>$  indicate  the average over the N particles, 
and we fit the value of the exponent $\alpha$ 
in eq. (\ref{anoma}). 
In fig.1 we plot on a log-log scale $\sigma_{\theta}^2$ vs. $t$ 
for $N=500$ at three different energies: U=0.2, U=0.6, U=5. 
The continuous lines are shifted fits and show a very clear power law
over a few decades; the corresponding values for the slope $\alpha$ 
are indicated in the inset. 
The numerical results show clearly three different types of behavior:  
  
1) No diffusion for very low energy, i.e. $U \le 0.2$.  
In this case all the particles belong to a single 
cluster and $\alpha=0$.

2) A ballistic regime $\alpha=2$ for $U$ bigger than the critical 
energy $(U_c=0.75)$ (a short-time ballistic regime is obviously
always present for all energies).

3) Superdiffusion with $\alpha = 1.38 \pm 0.05$ for U=0.6,  
in the transient regime. After a crossover time 
$\tau_c \sim 7 \cdot 10^4$ a change to the slope $\alpha = 1$ (normal 
diffusion) is observed. 
The superdiffusive regime is present in the energy range 
$0.5<U<0.75$.

In fig.2 we study the dependence on $N$ of the anomalous diffusion 
and the coincidence of crossover time $\tau_c$ with the relaxation time $\tau_r$,
i.e. the time the system needs to reach the canonical
temperature (horizontal dotted line in panel b)). 
We report $\sigma_{\theta}^2$ and temperature vs. time
for $N=500,2000$ and $U=0.69$. 
A slope $\alpha=1.42 \pm 0.05$ is observed in a first time stage, in
which the temperature is different from the canonical value.
In fact the temperature  maintains  for a very long period 
a constant value
which corresponds to a QSS belonging to  the continuation of the 
 homogeneous phase at a subcritical energy (see in particular 
fig.1 of Refs.\cite{latora}, and Ref.\cite{ruffo}).
Indeed, the crossover time from anomalous  to normal diffusion
$\tau_c$  coincides  with the relaxation 
time $\tau_r$. This result  has also been checked, changing
the accuracy of the numerical simulation.
The transient regime, in which QSS and anomalous 
diffusion are present, increases linearly with 
N~\cite{latora}, consequently    
for N=2000 one gets superdiffusion over almost 3 decades. 
On the other hand, the slope $\alpha$ does not seem to
strongly depend on $N$ and moreover in the range $0.6< U< 0.69$
it varies from 1.38 to 1.42.
A similar scenario has been recently conjectured by Tsallis \cite{tsallis1}
for systems with long-term memory and slow relaxation to equilibrium,
but to our knowledge this is the first time that it 
has been found in a numerical simulation.

The importance of noise and finite-size fluctuations 
 in the crossover from anomalous 
to normal diffusion has been studied in detail in 
Refs.\cite{grigo,barkai,torc}.
%
On the contrary, Kaneko and Konishi \cite{kaneko} claim that relaxation
to normal diffusion is due to phase-space uniform sampling, which
occurs asimptotically.
Our results show that this relaxation occurs in coincidence
with relaxation to equilibrium of QSS, which is a quite close
mechanism to the one proposed in Ref.\cite{kaneko}.
However, at variance with these latter authors, our model
displays superdiffusion, rather than subdiffusion, in the
transient.
Some important physical facts might be crucial in the observed
differences. Superdiffusion occurs near a second-order phase
transition in our case, while in the model of Ref.\cite{kaneko}
no phase transition is present. The model of Ref.~\cite{torc}
has a first-order phase transition and the particles performing
correlated flights belong only to a distinct dynamical class 
or phase for $N\to \infty$. In our case this is not true: 
close to the critical energy fluctuations are maximal and 
do not disappear in the thermodynamical limit. Each particle 
 regularly 
performs free walks and trapped oscillations until it forgets 
the initial condition and tends to a brownian motion.

Evidence in favour of  this mechanism is provided by the  link 
of superdiffusion with L\'evy walks. 
For low-dimensional chaotic Hamiltonian systems, 
superdiffusion has been interpreted  
as due to the trapping of the particles 
by the cantori of the phase space;  
particles can eventually escape and walk freely before 
a new trapping occurs \cite{levypro}, and 
this mechanism prevents from normal diffusion.   
An analogous   situation occurs in the experiment of Solomon et al.
for chaotic transport in a 
two-dimensional rotating flow \cite{sol}. In this case 
tracer particles are trapped and untrapped by a chain of 
six vortices.   
This last mechanism is very similar to  
ours. In fig.3 we report the time behavior 
of the angle $\theta$ and of the corresponding conjugate 
momentum $p$ of a test particle in the transient anomalous 
diffusion regime (panels (a) and (b)) and in the 
equilibrium regime (panels (c) and (d)) for $U=0.6$ and $N=500$.
Free walks and trapped motion are observed in the transient regime; 
the walks have an almost constant velocity corresponding 
to the separatrix between bounded and free motion ($\sim 2\sqrt{M}$) of the
perturbed pendulum of Eqs.~(\ref{eqmoto}).
In the equilibrium regime the test particle remains trapped in the cluster;
it oscillates around its center and drifts together with it 
on a much longer time scale (for a study of 
cluster motion see \cite{antoni}).
 It is important to notice that the energy of the test particle is not conserved.
The particle walks freely when accidentally it receives enough energy
that allows it to escape from the mean field. 
In this sense  the mechanism of anomalous 
diffusion in our case is similar to that of non-conservative  systems.
%
A quantitative difference between the two behaviors can be obtained by 
performing the power spectrum of the motion in fig.3(a) and 3 (c).
We get a power law with slope $-2$ for the equilibrium regime, as it 
should be for brownian motion, and a slope smaller than $-2$
for the transient.
To study the connection between L\'evy walks and anomalous
diffusion we evaluate trapping and walking time distributions. 
A free walk is identified by $\Delta \theta~ > ~2 \pi$.  
In fig.4 we consider for $N=500$ and two energies U=0.6 and U=0.69, 
the probability distribution of ``walking times'' and ``trapping times''.
They show, as expected, a clear power law decay:  
\begin{equation}
    P_{walk}   \sim  t^{-\mu}, ~~~~~~~
    P_{trap}    \sim  t^{-\nu}~~.
\end{equation}
The values of $\mu$ and $\nu$ obtained from the fitting are reported in 
figure. Their value is crucial because,  
the two exponents $\mu$ and $\nu$ can 
be related to the anomalous diffusion coefficient $\alpha$. 
The following relationships, derived  in ref. \cite{kla},
are the most 
appropriate for HMF: 
\begin{eqnarray}
\alpha= 2 + \nu -\mu  ~~~~~~~ 2<\mu<3, ~~\nu<2   
\\
\alpha= 4 -\mu        ~~~~~~~~~~~ 2<\mu<3, ~~\nu>2 .
\end{eqnarray}
These formulae are valid for a one-dimensional system  
under the assumptions of walks with a constant velocity, 
separated by sticking events with no motion. 
In the case shown we get for U=0.69 ~~ (0.6) 
$\mu=2.14\pm 0.1 ~~~ (1.98\pm 0.1)$, $\nu=1.58 \pm 0.05 ~~~(1.34 \pm 0.05)$ 
and thus a value for $\alpha=2+\nu-\mu=1.44 \pm 0.05 ~~ (1.36 \pm 0.05)$ 
which is consistent with
the values obtained from the fits 
shown in figs.1 and 2, 
within our
numerical accuracy (the relative error on the exponent $\alpha$
obtained by fitting the slope of the variance is $\sim 5\%$). 
As found in ref.\cite{sol} we get for the trapping probabilities the
exponent 
$\nu <2 $, which is not the usual case encountered in low dimensional
conservative 
maps. The reason of that is likely  the non-conservation of 
energy  for  the test particle motion.
%
%
In conclusion, we have found superdiffusion and L\'evy walks in 
a Hamiltonian system showing a second-order phase transition.
This behavior occurs in a transient out-of-equilibrium regime
in a range of energies sligthly smaller than the critical energy
where  the system is strongly chaotic and QQS exist. 
We have also found that the equilibration time to reach the 
canonical temperature, which diverges with N,
corresponds to the crossover  to normal diffusion confirming a 
recently proposed scenario \cite{tsallis1}.
This feature has been observed for the first time  in a deterministic
chaotic system with many degrees 
of freedom and could be of relevance to understand
more realistic situations such as the anomalous diffusion observed 
in fluid flow experiments \cite{sol}. 

We thank 
M. Antoni, M. Baranger, E. Barkai, A. Torcini and C. Tsallis  
%
for  stimulating discussions. 
A.R. thanks the Center for Theoretical Physics of MIT for the kind 
hospitality. V.L. thanks CNR and Blanceflor-Ludovisi Foundation for 
financial support. 
S.R. thanks the Laboratoire de Physique at ENS Lyon for hospitality
and CNRS-URA 1325 for financial support.

(1) Electronic adress: Latora@ctp.mit.edu

(2) Electronic adress: Rapisarda@ct.infn.it

(3) Electronic adress: Ruffo@avanzi.de.unifi.it

\begin{figure}
\begin{center}
\epsfig{figure=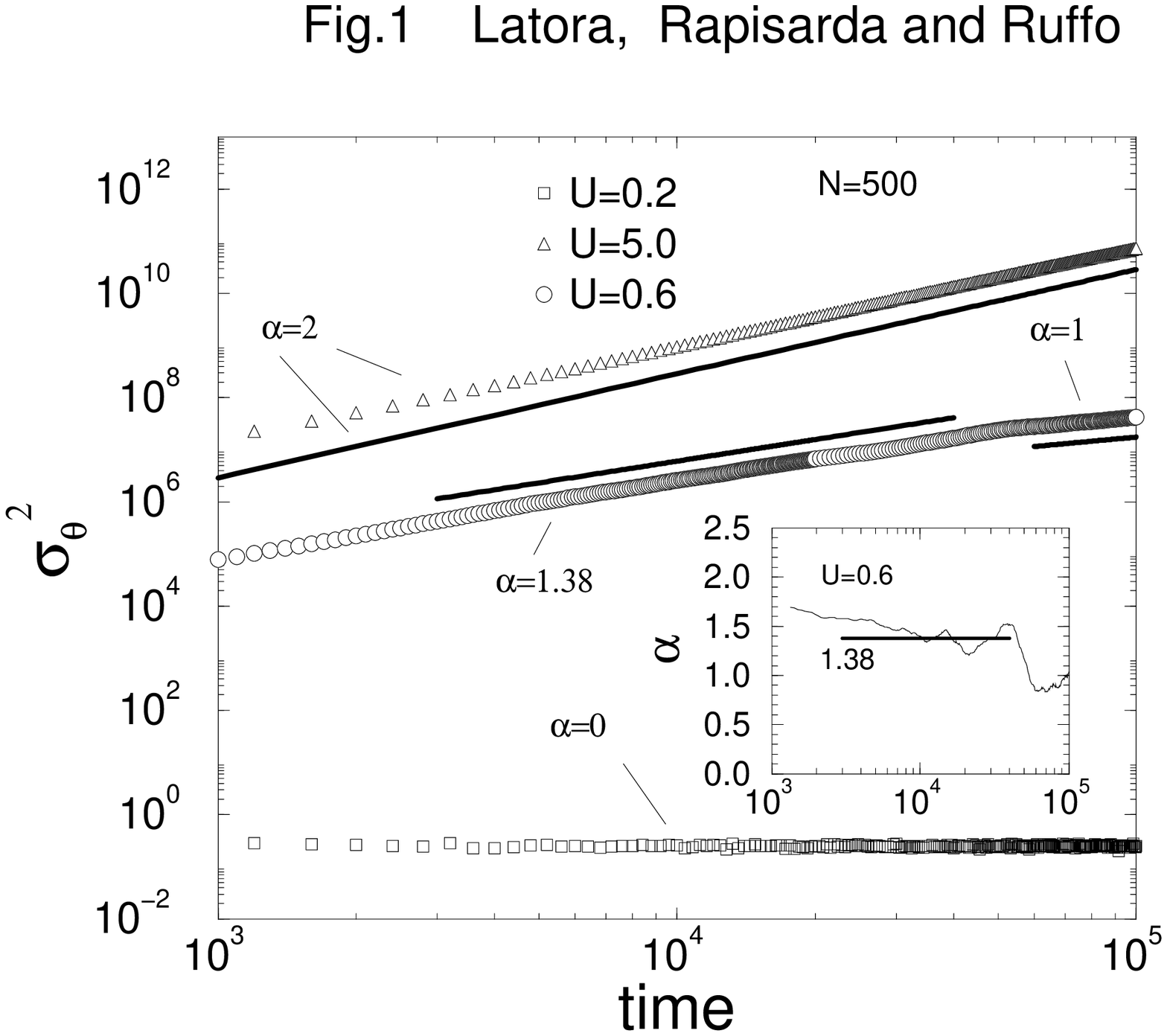,width=\columnwidth,angle=0}
\end{center}
\caption{{\small For N=500 we show three different behaviors: 
1) no diffusion for U=0.2, 
2) ballistic diffusion for U=5 and 
3) superdiffusion for U=0.6. 
In this last case we considered the average over 5 events. 
The straight full lines are  shifted fits and   
the relative slopes are also indicated. The relative errors
obtained from the fits are $\sim 5 \%$. 
We show in the inset the numerical evaluation of the slope $\alpha$ vs time
for the case U=0.6. }}
\label{fig:diffu1}
\end{figure}
\begin{figure}
\begin{center}
\epsfig{figure=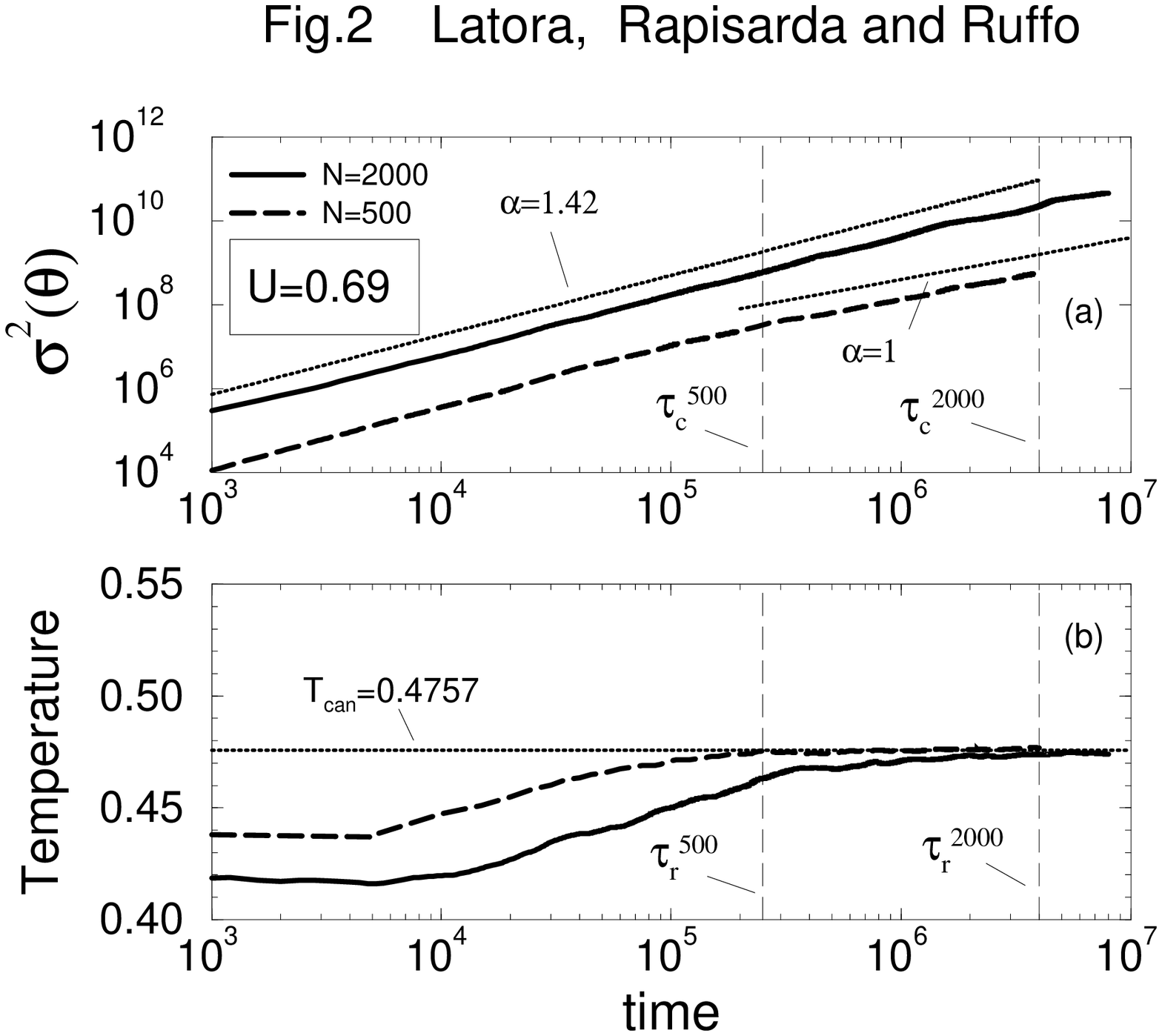,width=\columnwidth,angle=0}
\end{center}
\caption{\small Variance and temperature for
two different sizes N=500 (dashed lines),2000 (full lines) at U=0.69. 
Panel (a) shows that $\alpha \sim 1.4$ does not depend on N 
(within the accuracy of the calculations) and occurs only in a 
transient regime. 
Once the canonical temperature, shown in panel (b), is reached,  
diffusion becomes normal. 
The relaxation time $\tau_r$ is clearly larger for bigger $N$.
The vertical dashed lines indicate $\tau_c \sim \tau_r$
for N=500 and N=2000.}
\label{fig:diffu2}
\end{figure}
\newpage
\begin{figure}
\begin{center}
\epsfig{figure=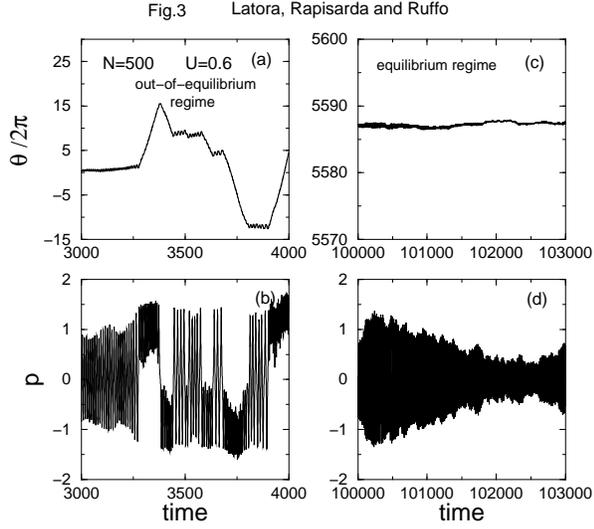,width=\columnwidth,angle=0}
\end{center}
\caption{\small Angle $\theta$ and momentum $p$ of a typical particle
for $U=0.6$ and $N=500$. Panels (a) and (b) refer to the transient
regime, (c) and (d) to the canonical equilibrium state.
See text for more details.}
\label{fig:diffu3}
\end{figure}
\begin{figure}
\begin{center}
\epsfig{figure=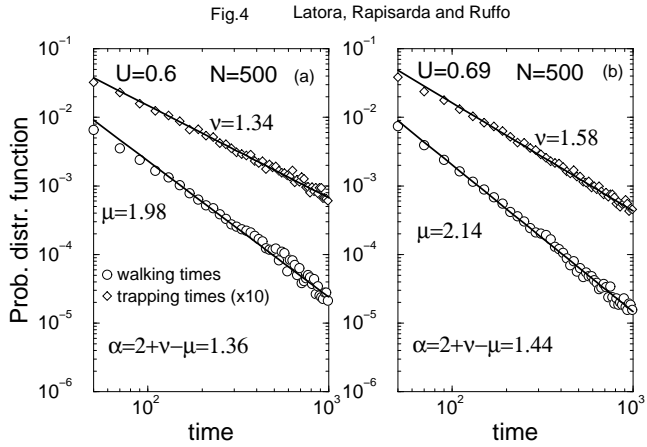,width=\columnwidth,angle=0}
\end{center}
\caption{\small For the cases U=0.6 and U=0.69, we show the probability 
distribution functions for trapping (open diamonds) and walking times
(open circles) calculated in the transient regime ($2000<t<8000$). 
The correspondent fits and exponents are 
also indicated, see text for more details. }
\label{fig:diffu4}
\end{figure}

\vfill

\end{document}